# Extended Chiral Quark Models in the Framework of Quantum Chromodynamic: Theory and their Applications in Hot and Dense Mediums

M. Abu-Shady

*Department of Applied Mathematics, Faculty of Science, Menoufia University, Egypt.*

**Corresponding Author:** *M. Abu-Shady, dr.abushady@gmail.com*

**ABSTRACT**

*In this review article, we give overview on the extended chiral quark models. In particular, how these models are extended to include the higher-order interactions, quantized fields, logarithmic potential and the effect of these modifications on hadron properties at hot and dense mediums. In addition, how we deal with non-normalization of the Nambu-Jona-Lasinio (NJL) model by using mid-point technique. Therefore, the extended quark models give satisfied results in comparison with lattice QCD at finite temperature and baryonic chemical potential. The extra effort is needed to deal with chiral quark models in the presence of external magnetic field at finite temperature and density.*

## INTRODUCTION

Investigation of hadron properties unfortunately cannot be calculated directly in the framework of QCD theory [1, 2]. The difficulties involved in obtaining low-energy properties directly from QCD, the fundamental theory of strong interactions is due to its coupling constant at low energy scale and thereby the necessity of dealing non-perturbatively with its complicated structure. Hence, the investigations are usually performed by using the effective models which share the same properties with QCD theory such as the Nambu-Jona-Lasinio (NJL) model [3] and the linear sigma model and its modifications [4-15] at zero temperature and chemical potential. The linear sigma model has been proposed as a model for strong nuclear interactions [4]. The model was first proposed in the 1960s as a model for pion-nucleon interactions. The model exhibits the spontaneous breaking of chiral symmetry and its restoration at finite temperatures. At finite temperature, the model gives a good description of the phase transition by using the Hartree approximation [16, 17] within the Cornwall--Jackiw--Tomboulis (CJT) formalism [18].

At finite temperature and density, the linear sigma model and its modifications show a significant success in the description of chiral phase transition and the static properties of the nucleon such as discussed in Refs. [19-26]. There are some limited research works on the nucleon properties at finite temperature and density in the framework of the linear sigma model. Christov et al. [19] studied the modification of baryon properties at finite temperature and density. They ignored quantum fluctuations in their calculations. Dominguez et al. [20] calculated the pion-nucleon coupling constant and the mean square radius of the proton as functions of the temperature. Thermal fluctuations are only considered and they ignored quantum fluctuations in their model. In Ref. [21], the nucleon properties are also studied in the framework of the linear sigma model at finite temperature without including chemical potential in their calculations.

On the other hand, in Refs. [22-24], the authors focus on the study of the phase transition and critical point at finite temperature and chemical potential in the framework of the linear sigma model. Bilic and Nikolic [22] studied the chiral transition in the linear sigma model at finite temperature and chemical potential in mean-field approximation. Phat and Thu [25] studied the chiral phase transition and its order at finite temperature and isospin chemical potential using the linear sigma model. Schaefer et al. [26] have extended the quark sigma model to include certain aspects of the gluon dynamics via the Polyakov loop at finite temperature and quark chemical potential. Mao et al. [27] studied the deconfinment phase transition using the Friedberg-Lee model at finite temperature and chemical potential, and then they obtained the critical value of the temperature and chemical potential. In the NJL model [28], the meson and nucleon properties are studied. they considered the case of a quark medium as well as nucleon medium. They studied the behavior of the nucleon mass and energy of the soliton at different temperatures and densities in the mean-field approximation. Cheng-Fu et al. [29] studied some nucleon properties such as a binding energy using the simplified version of Faddeev equations at finite temperature and chemical potential.

In this paper, we review the recent extended quark models. In Sec. (2), we give overview on the higher-order quark model, the logarithmic quark model, the quantized quark model, and the NJL model which play an important role at hot and dense mediums. The summary and conclusion are given in Sec. (3).

## EXTENDED CHIRAL QUARK MODELS

In the recent years, the multi-quark interactions have played an important role in studying the phase transition in the chiral quark models. The Nambu-Jano-Lasinio (NJL) model has been extended to higher-order quark interactions in the presence the finite temperature and chemical potential. The effect of these interactions on the phase transition has been investigated [30, 31]. In Ref. [32], the effect of the higher-order mesonic interactions are studied on nucleon properties such as the nucleon mass, the magnetic moments of the proton and neutron, and the pion-nucleon coupling constant at finite temperature in the linear sigma model. The Lagrangian density of the extended sigma model which describes the interactions between quarks via the exchange of σ- and π-mesons at finite temperature is written

$$L(r) = i\overline{\Psi}\gamma^\mu \partial_\mu \Psi + \frac{1}{2}\left(\partial_\mu \sigma \partial^\mu \sigma + \partial_\mu \boldsymbol{\pi} \cdot \partial^\mu \boldsymbol{\pi}\right) + g\overline{\Psi}(\sigma + i\gamma_5 \boldsymbol{\tau} \cdot \boldsymbol{\pi})\Psi - U_{eff}^{T(1)}(\sigma, \boldsymbol{\pi}), \quad (1)$$

where,

$$U_{eff}^{T(1)} = U_1^{T(0)}(\sigma, \boldsymbol{\pi}) + \frac{7\pi^2 T^4}{90} + \left(\frac{m_\sigma^2 - m_\pi^2}{24 f_\pi^2}\right)T^2\left(\sigma^2 + \boldsymbol{\pi}^2\right) +$$

$$\left(\frac{m_\sigma^2 - m_\pi^2}{24 f_\pi^2}\right)T^2\left(\sigma^2 + \boldsymbol{\pi}^2 - \frac{v^2}{2}\right), \quad (2)$$

with

$$U_1^{T(0)}(\sigma, \boldsymbol{\pi}) = \frac{\lambda^2}{4}\left(\sigma^2 + \boldsymbol{\pi}^2 - v^2\right)^2 + m_\pi^2 f_\pi \sigma. \quad (3)$$

$U_1^{T(0)}(\sigma, \boldsymbol{\pi})$ is the original meson-meson interaction potential at zero temperature, where $\Psi, \sigma,$ and $\boldsymbol{\pi}$ are the quark, sigma, and pion fields, respectively. In the mean-field approximation, the meson fields are treated as time-independent classical fields. This means that we replace the powers and products of the meson fields by their corresponding powers and the products of their expectation values. The meson-meson interactions in Eq. (3) lead to hidden chiral $SU(2) \times SU(2)$ symmetry with $\sigma(r)$ taking on the vacuum expectation value

$$\langle \sigma \rangle = f_\pi, \quad (4)$$

where $f_\pi = 93$ MeV is the pion decay constant. The final term in Eq. (3) is included to break the chiral symmetry and to generate the quark mass. It leads to the partial conservation of the axial-vector isospin current (PCAC). The parameters $\lambda^2, v^2$ can be expressed in terms of $f_\pi$, the masses of mesons

$$\lambda^2 = \frac{m_\sigma^2 - m_\pi^2}{2 f_\pi^2}, \quad (5)$$

and

$$v^2 = f_\pi^2 - \frac{m_\pi^2}{\lambda^2}. \quad (6)$$

# Extended Chiral Quark Models in the Framework of Quantum Chromodynamic: Theory and their Applications in Hot and Dense Mediums

We construct the effective potential with higher-order mesonic interactions in the linear sigma model. The temperature-dependent effective potential in the one-loop approximation is given by:

$$U_{eff}^{T(2)} = U_2^{T(0)}(\sigma, \pi) + \frac{7\pi^2 T^4}{90} + (\frac{m_\sigma^2 - m_\pi^2}{24 f_\pi^2})T^2(\sigma^2 + \pi^2) +$$

$$(\frac{m_\sigma^2 - m_\pi^2}{24 f_\pi^2})T^2\left(\sigma^2 + \pi^2 - \frac{v^2}{2}\right). \quad (7)$$

To include higher-order mesonic interactions, we assume the meson potential $U_2^{T(0)}$ at zero temperature is as the same form in [11]:

$$U_2^{T(0)}(\sigma, \pi) = \frac{\lambda^2}{4} A\left((\sigma^2 + \pi^2)^2 - Bv^2\right)^2 + m_\pi^2 f_\pi \sigma, \quad (8)$$

where the potential satisfies chiral symmetry and has eight-point interactions. Applying the minimizing conditions and the PCAC, we get

$$A = \frac{m_\sigma^2 - 3m_\pi^2}{4 f_\pi^4 (m_\sigma^2 - m_\pi^2)} \quad (9)$$

and

$$B = f_\pi^2 \left(1 - \frac{2m_\pi^2(m_\sigma^2 + m_\pi^2)}{(m_\sigma^2 - 3m_\pi^2)^2}\right) \quad (10).$$

By using the Euler-Lagrange equations, the meson field equations are derived

$$\nabla^2 \sigma' = g\overline{\Psi}\Psi - 2A\lambda^2(\sigma' - f_\pi)\left((\sigma' - f_\pi)^2 + \pi^2\right)\left(\left((\sigma' - f_\pi)^2 + \pi^2\right)^2 - Bv^2\right) - m_\pi^2 f_\pi$$
$$-(\frac{m_\sigma^2 - m_\pi^2}{6 f_\pi^2})T^2(\sigma' - f_\pi) \quad (11)$$

and,

$$\nabla^2 \pi = ig\overline{\Psi}\gamma_5 \tau \Psi - 2A\lambda^2 \pi\left((\sigma' - f_\pi)^2 + \pi^2\right)\left(\left((\sigma' - f_\pi)^2 + \pi^2\right)^2 - Bv^2\right)$$
$$-(\frac{m_\sigma^2 - m_\pi^2}{6 f_\pi^2})T^2 \pi, \quad (12)$$

Where $\tau$ refers to Pauli isospin-matrices, $\gamma_5 = \begin{pmatrix} 0 & 1 \\ 1 & 0 \end{pmatrix}$.

We used the hedgehog ansatz [5] where

$$\pi(r) = \pi(r)\hat{\mathbf{r}}. \quad (13)$$

By using the Dirac equation, the quark field equations are derived

$$\frac{du}{dr} = -P(r)u + (W - m_q + S(r))w, \quad (14)$$

where, $S(r) = g\langle \sigma' \rangle$, $P(r) = \langle \pi.\hat{\mathbf{r}} \rangle$ and $W$ are the scalar potential, the pseudoscalar potential and the eigenvalue of the quarks spinor $\Psi$, respectively.

$$\frac{dw}{dr} = -(W - m_q + S(r))u + \left(\frac{2}{r} - p(r)\right)w. \quad (15)$$

then, we used an iteration method by employing the Green's functions for solving the above system of equations (11,12, 13, 14). The total energy of the nucleon is calculated, which consists of quark, sigma, pion, quark-sigma interaction, quark-pion interaction, and meson static energy contributions. The nucleon mass was derived as in Ref. [11]. One of these results is displayed in Fig. (1).

**Extended Chiral Quark Models in the Framework of Quantum Chromodynamic: Theory and their Applications in Hot and Dense Mediums**

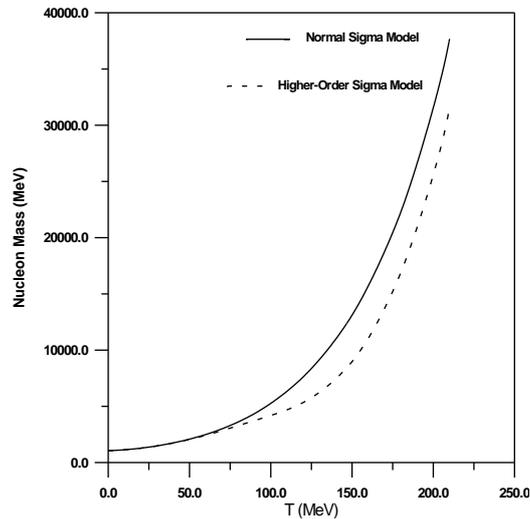

**Fig. 1.** *The nucleon mass is plotted as a function of the temperature T for both the original sigma model and the higher-order sigma model.*

In Fig. (1), the nucleon mass is plotted as a function of the temperature $T$ for the higher-order sigma model and the original sigma model. We note that the two curves show similar behavior. In the higher-order sigma model, the nucleon mass starts at 1040 MeV which is slightly less than the $M_N = 1049$ MeV in the original sigma model at zero temperature. We note that the nucleon mass changes only slightly with increasing temperature up to 60 MeV in the two models. Then, the effect of the higher-order mesonic interactions clearly appears at higher values of the temperature. We find that the nucleon mass assumes lower values in the higher-order sigma model compared to the original sigma model. This suggests that the coupling between the quark and meson field increases with increasing higher-order mesonic interactions in the original sigma model, therefore a reduction in nucleon mass is obtained.

**Table 1.** *The energy calculations of the nucleon mass for two models: The original sigma model and the higher-order sigma model for $g = 4.94, m_\pi = 139.6\,MeV, m_\sigma = 600\,MeV$, $f_\pi = 93$ MeV, and $T = 120$ MeV. All quantities are expressed in MeV.*

| Quantity | Original sigma model | Higher-order sigma model |
|---|---|---|
| Quark kinetic energy | 781.197 | 462.973 |
| Sigma kinetic energy | 394.767 | 769.682 |
| Pion kinetic energy | 203.661 | 25.533 |
| Sigma interaction energy | 15.313 | 121.339 |
| Pion interaction energy | -412.807 | -106.192 |
| Meson interaction energy | 6637.39 | 4065.78 |
| Hedgehog mass baryon | 7619.49 | 5339.1 |
| Nucleon mass | 7610.63 | 5334.42 |

In Table (1), we show the dynamics of the nucleon mass in the normal sigma model and the higher-order sigma model at a fixed value of the temperature T=120 MeV. We observe that the quark and pion kinetic energies decrease with increasing mesonic interactions in the higher-order sigma model. The sigma kinetic energy and sigma interaction increase with increasing mesonic contributions. Moreover, the meson-meson interaction reduces with increasing mesonic interactions in the normal sigma model. This indicates that the coupling between mesons is more tight in the higher-order sigma model at higher-values of the temperature. Therefore, a stong reduction in nucleon mass is obtained for the higher-order sigma model. Another attempt, by including quantized of fields at finite temperature. The quantization of fields are considered only at zero temperature such as in Refs. [8,9].

In Ref. [33], the authors suggest quantized of field at finite temperature. Therefore, the Hamiltonian density of the linear sigma model that describes the interactions between quarks

**Extended Chiral Quark Models in the Framework of Quantum Chromodynamic: Theory and their Applications in Hot and Dense Mediums**

via the σ-and π-mesons is written as [9]

$$H(r) = \frac{1}{2}\{\hat{P}_\sigma(r)^2 + (\nabla\hat{\sigma}(r))^2 + \hat{P}_\pi(r)^2 + (\nabla\pi(r))^2\} + U^{eff}(\hat{\sigma},\hat{\pi})$$

$$\overline{\Psi}(r)(-i\alpha\cdot\nabla)\hat{\Psi}(r) - g\overline{\Psi}\beta(r)(\hat{\sigma}(r) + i\gamma_5\hat{\tau}\cdot\hat{\pi})\hat{\Psi}(r) \quad (16)$$

with

$$U^{eff}(\hat{\sigma},\hat{\pi}) = \frac{\lambda^2}{4}(\hat{\sigma}^2 + \hat{\pi}^2 - v^2)^2 - f_\pi m_\pi^2 \hat{\sigma} - \frac{\pi^2}{45}T^4 + \frac{T^2}{24 f_\pi^2}(3m_\sigma^2 - 5m_\pi^2)(\hat{\sigma}^2 + \hat{\pi}^2), \quad (17)$$

$$\lambda^2 = \frac{m_\sigma^2 - m_\pi^2}{2f_\pi^2}, \quad v^2 = f_\pi^2 - \frac{m_\pi^2}{\lambda^2} \quad (18)$$

In the Eq. (17), $\hat{\Psi}, \hat{\sigma}$, and $\hat{\pi}$ are quantized field operators with the appropriate static angular momentum expansion [9]

By using variational method, we obtained the following linear and non-linear differential equations of quark and meson fields:

$$\frac{du}{dr} = -2(g\sigma + \varepsilon)v(r) - \frac{1}{3}\alpha\delta(a+b)g\Phi(r)u(r), \quad (19)$$

$$\frac{dv}{dr} = -\frac{2}{r}v(r) - 2(g\sigma(r) - \varepsilon)u(r) + \frac{1}{3}\alpha\delta(a+b)g\Phi(r)u(r), \quad (20)$$

$$\frac{d^2\sigma}{dr^2} = -\frac{2}{r}\frac{d\sigma}{dr} - m_\pi^2 f_\pi + 3g(u^2(r) - v^2(r)) + \lambda^2(N_\pi + x)(\sigma^2(r) - v^2)\Phi^2(r)\sigma(r) +$$

$$+\lambda^2(\sigma^2(r) - v^2)\sigma(r) + \frac{T^2}{12 f_\pi^2}(3m_\sigma^2 - 5m_\pi^2)\hat{\sigma}, \quad (21)$$

$$\frac{d^2\Phi}{dr^2} = -\frac{2}{r}\frac{d\Phi}{dr} + \frac{2}{r^2}\Phi(r) + \frac{1}{2}(1 - \frac{x}{N_\pi})m_\pi^2\Phi^2 + \frac{\lambda^2}{2}(1 + \frac{x}{N_\pi})(\sigma^2(r) - v^2)\Phi(r) -$$

$$\frac{\alpha}{N_\pi}(a+b)g\delta u(r)v(r) + \frac{\lambda^2}{N_\pi}[x^2 + 2xN_\pi + 81(\alpha^2 a^2 c^2 + (\beta^2 + \gamma^2)d^2)]\Phi^3(r)$$

$$-\frac{k}{N_\pi}\Phi_p(r) + \frac{T^2}{24 f_\pi^2}(3m_\sigma^2 - 5m_\pi^2)\Phi(r)\left(1 + \frac{x}{N_\pi}\right), \quad (22)$$

where a and b are functions in the coherent parameter $x$ and $\alpha^2 + \beta^2 + \gamma^2 = 1$. The $\sigma$, $\Phi$, u, and v are sigma, pion, and components of quark fields, respectively. One of the results of this work is displayed in Table (3).

**Table3.** The observables of the nucleon calculated for two values of coherence parameter, $x = 0.3$ and $x = 3$ at $T_c = 161$ MeV, $m_\sigma = 450$ MeV and $g = 4.5$

| $x$ | $x = 0.3$ | | | $x = 3$ | | |
|---|---|---|---|---|---|---|
| Quantity | Quark | Meson | Total | Quark | Meson | Total |
| $\langle r^2 \rangle_p$ | 3.937 | $-2.370\times10^{-5}$ | 3.937 | 4.655 | $4.75\times10^{-3}$ | 4.660 |
| $\langle r^2 \rangle_n$ | $-6.171\times10^{-5}$ | $2.370\times10^{-5}$ | $-3.8\times10^{-5}$ | $6.77\times10^{-3}$ | $-4.753\times10^{-3}$ | $2.01\times10^{-3}$ |
| $\mu_p$ | 2.4928 | $1.099\times10^{-5}$ | 2.492 | 1.541 | $1.358\times10^{-2}$ | 1.555 |
| $\mu_n$ | -1.661 | $-1.099\times10^{-5}$ | -1.661 | -1.031 | $-1.358\times10^{-2}$ | -1.0445 |
| $g_A(0)$ | 0.631 | $1.377\times10^{-5}$ | 0.631 | 0.4696 | $1.448\times10^{-2}$ | 0.484 |
| $g_{\pi NN}$ | 1.3898 | $7.6\times10^{-4}$ | 1.390 | 0.864 | -0.105 | 0.7596 |

Table (2) shows that the nucleon properties are sensitive to the change of the coherence parameter $x$. The mesonic contributions are sensitively affected by changing the coherence parameter $x$ at the critical point temperature where the mean-square radius of the proton is increased by about 15%. Also, the mean-square radius of the neutron is observed to strongly change from a negative value to a positive value. We note that the mesonic contributions in the charge radius of the proton and neutron increase when the coherence parameter $x$ is increased. A similar situation is found for the magnetic moment of the proton and neutron, in which the mesonic contributions increase with increasing parameter $x$. In addition, the $g_A(0)$ values are reduced by about 23% with increasing $x$. The $g_{\pi NN}(0)$ strongly decreases with increasing $x$, since the coupling constant is weakened by the increase of $x$.

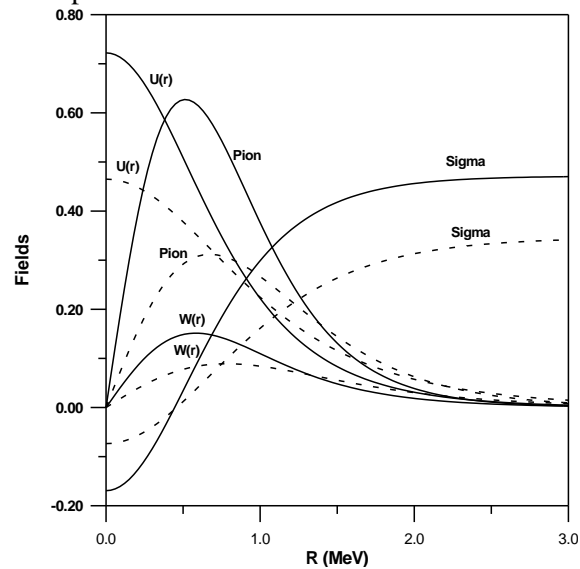

**Fig. 2.** *Sigma, pion and components of quark fields are plotted as functions of the distance R where the bold curves are for T = 0 and the dashed curves are for T = 100 MeV.*

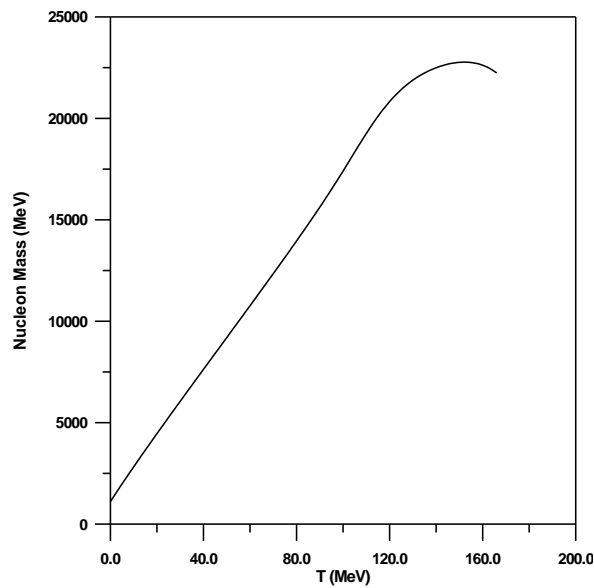

**Fig. 3.** *The nucleon mass is plotted as a function of the temperature T.*

From Fig.(2), we have plotted the above fields as functions of the distance r at zero and finite temperature (T=100 MeV), respectively. We note the behavior is similar in the two cases where the fields are shifted to lower values in comparison with the zero temperature case. In Fig. (3), the nucleon mass is plotted as a function of the temperature. One can see that the nucleon mass monotonically increases with increasing temperature and slightly decreases at higher values of the temperature. They noticed that the nucleon mass increases until a value of $\frac{3}{4}T_C$ ($T_C$=161 MeV is critical temperature).

Dominguez and Loewe [20] deduced that the nucleon mass increases with increasing temperature. They studied the nucleon propagator at finite temperature in the framework of finite energy QCD sum rules. They also interpreted their results as a phase deconfinment transition. This work is in agreement with the Dominguez and Loewe [20] calculation

In Ref. [21], the coherent-pair approximation (CPA) is applied in the chiral quark-sigma model by fully taking thermal and quantum fluctuations into account. So far no attempt has been made to include the finite chemical potential. We can write the Hamiltonian of quark sigma model at finite temperature (T) and chemical potential (μ) including quantized fields

$$\hat{H}(r) = \frac{1}{2}\{\hat{P}_\sigma(r)^2 + (\nabla\hat{\sigma}(r))^2 + \hat{P}_\pi(r)^2 + (\nabla\hat{\pi}(r))^2\} + U_2^{eff}(\hat{\sigma}, \hat{\pi}) +$$

$$\overline{\hat{\Psi}}(r)(-i\alpha\nabla)\hat{\Psi}(r) - g(r)\overline{\hat{\Psi}}(r)(\beta\hat{\sigma}(r) + i\beta\gamma_5\hat{\tau}\cdot\hat{\pi})\hat{\Psi}(r), \qquad (23)$$

the effective potential takes the following form in the one-loop approximation.

$$U_2^{eff}(\sigma, \pi) = U^{(0)}(\sigma, \pi) - \frac{6}{\pi^2}\left(\frac{7\pi^4}{180}T^4 + \frac{\pi^2}{6}T^2\mu^2 + \frac{1}{12}\mu^4\right) +$$

$$6g^2\left(\frac{T^2}{12} + \frac{\mu^2}{4\pi^2}\right)(\sigma^2 + \pi^2). \qquad (24)$$

where, the original mesonic potential at zero temperature and chemical potential is given by

$$U^{(0)}(\sigma, \pi) = \frac{\lambda^2}{4}(\sigma^2 + \pi^2 - \nu^2)^2 + m_\pi^2 f_\pi \sigma, \qquad (25)$$

and third and fourth terms depend on the temperature and chemical potential, and quantized fields. The variational method is used to obtain the system of non-linear differential equations and the nucleon properties are obtained as in Table (3). For more discussion, see original work [33].

**Table 3.** *The observables of the nucleon calculated for a value of the coherence parameter $x = 1$ at $m_\pi = 139.6$ MeV, $m_\sigma = 472$ MeV and $g = 4.5$*

| Quantity | $T = 0$ and $\mu = 0$ | | | $T = 0$ and $\mu = 270$ MeV | | |
|---|---|---|---|---|---|---|
| | Quark | Meson | Total | Quark | Meson | Total |
| $\langle r^2 \rangle_p$ | $7.217\times10^{-1}$ | $2.8605\times10^{-2}$ | 0.750 | 2.535 | $-2.584\times10^{-5}$ | 2.535 |
| $\langle r^2 \rangle_n$ | $2.232\times10^{-2}$ | $-2.8605\times10^{-2}$ | $-6.277\times10^{-3}$ | $-9.187\times10^{-6}$ | $2.584\times10^{-5}$ | $1.665\times10^{-5}$ |
| $\mu_p$ | 1.734 | 0.175 | 1.909 | 0.750 | $6.659\times10^{-7}$ | 0.750 |
| $\mu_n$ | -1.262 | -0.175 | -1.437 | 0.500 | $6.659\times10^{-7}$ | -0.500 |
| $\frac{g_A}{g_V}$ | 1.1624 | 0.371 | 1.534 | 0.296 | $2.614\times10^{-7}$ | 0.296 |
| $g_{\pi NN}^G$ | 0.873 | 0.278 | 1.151 | 0.417 | $1.90\times10^{-7}$ | 0.417 |

The quark sigma model has some difficulties one of them that the critical temperature is not agreement with Lattice QCD. Therefore, we suggested the logarithmic quark model [34] to investigate the chiral phase transition. We give the logarithmic potential as follows

$$U_2^{(0)}(\sigma, \pi) = -\lambda_1^2(\sigma^2 + \pi^2) + \lambda_2^2(\sigma^2 + \pi^2)^2 \log\left(\frac{\sigma^2 + \pi^2}{f_\pi^2}\right) + m_\pi^2 f_\pi \sigma, \qquad (26)$$

this potential satisfies the chiral symmetry when $m_\pi \to 0$ as well as in the standard potential as in [5]. Spontaneous chiral-symmetry breaking gives a nonzero vacuum expectation for $\sigma$ and the explicit chiral-symmetry breaking term in Eq. 26 gives the pion its mass.

$$\langle\sigma\rangle = -f_\pi. \qquad (27)$$

By applying the minimized conditions, we obtain

**Extended Chiral Quark Models in the Framework of Quantum Chromodynamic: Theory and their Applications in Hot and Dense Mediums**

$$\lambda_1^2 = \frac{m_\sigma^2 - 7m_\pi^2}{12}, \tag{28}$$

$$\lambda_2^2 = \frac{m_\sigma^2 - m_\pi^2}{12 f_\pi^2}. \tag{29}$$

In Eq. 26, the potential has not a singularity point when we take limit $\sigma^2 + \pi^2 \to 0$. This advantage is not found Refs. [13, 14]. In Figs. $(4,5)$, the original and logarithmic potentials are plotted in the $(\sigma - \pi)$ plane. We note that the behavior is similar in the two potentials, but the logarithmic potential takes smaller values in comparison with the original potential. Therefore, the logarithmic potential is faster to restore the chiral symmetry by climbing the maximum located at the centre. Another advantage is that the logarithmic potential does not have any instability at any value of the sigma field. Hence, all these advantages will affected on the description on meson properties at low and high energies. As discussed in the above section, the critical temperature is reduced in comparison with original sigma model. Obtained value of critical temperature is in good agreement with lattice QCD results.

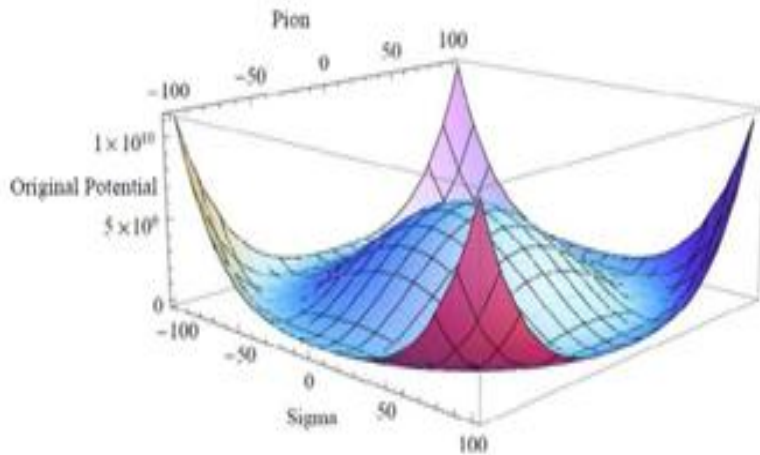

**Fig. 4.** *The original potential is plotted as a function of sigma and pion fields.*

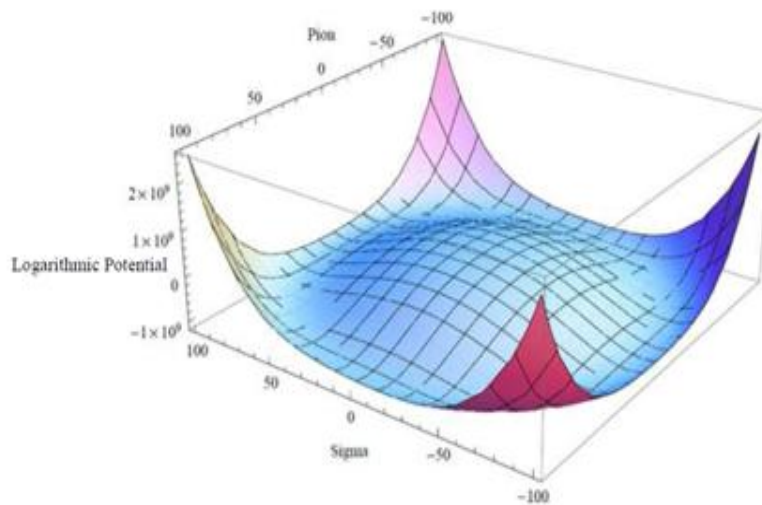

**Fig. 5.** *The logarithmic potential is plotted as a function of sigma and pion Fields*

Now, we can write the effective logarithmic potential describes the interactions of quarks via $\sigma$ and $\pi$-meson at finite temperature $T$ and baryonic potential $u_B$ as follows: The effective potential takes the following form in the one-loop approximation [23].

**Extended Chiral Quark Models in the Framework of Quantum Chromodynamic: Theory and their Applications in Hot and Dense Mediums**

$$U_2^{eff}(\sigma,\pi,T,u_B) = U_2^{T(0)}(\sigma,\pi) - \frac{6}{\pi^2}(\frac{7\pi^4}{180}T^4 + \frac{\pi^2}{6}T^2 u_B^2 + \frac{1}{12}u_B^4) +$$

$$6g^2(\frac{T^2}{12} + \frac{u_B^2}{4\pi^2})(\sigma^2 + \pi^2). \tag{30}$$

We give some results such as in Fig. 6 and Fig. 7.

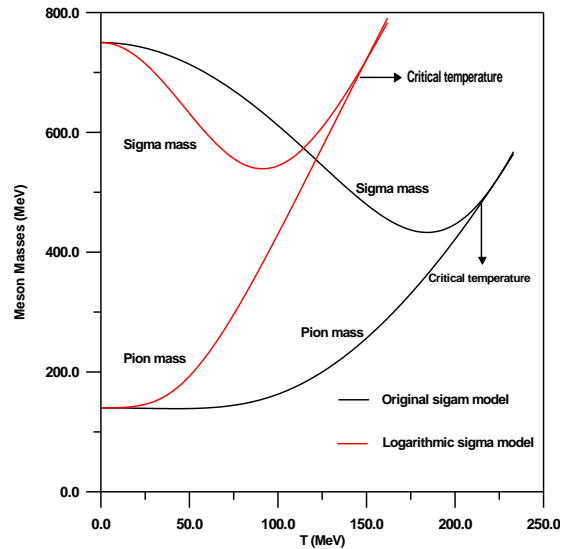

**Fig. 6.** *Sigma and pion masses are plotted as functions of temperature T in the above models in the presence of the explicit breaking symmetry term at $u_B = 0$. At temperature $T = 0$, the pions appear with the observed pion mass =140 MeV*

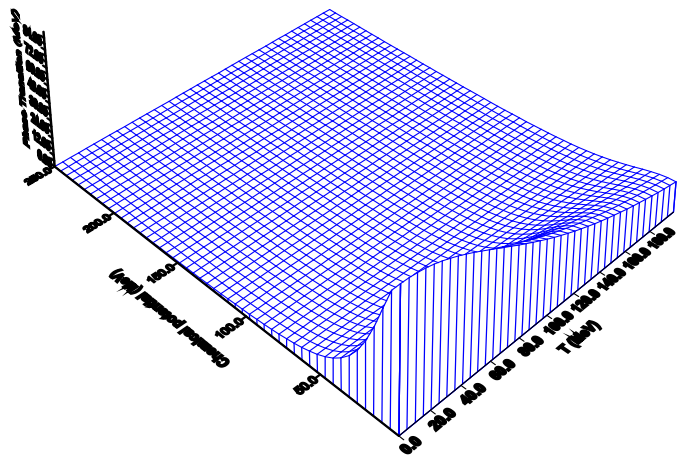

**Fig. 7.** *Phase transition is plotted as a function of temperature and chemical potential in the logarithmic sigma model.*

In Fig. (6), we show that the critical temperature is reduced in comparison with the original sigma model. This result is a good agreement with lattice QCD. In Fig. (7), the phase transition is crossover at finite temperature and zero chemical potential and also at zero temperature and finite chemical potential. We continue further study of the investigation. By studying how the nucleon properties respond to finite temperature and chemical potential in the framework logarithmic quark sigma model [35]. In following Table, we show the results of logarithmic quark model in comparison with original sigma model and its extended.

**Table 4.** *Values of the energy calculations for the hedgehog mass for the* Logarithmic Quark Model (LQM), *linear sigma model (LSM), and the extended sigma model (ESM) at* $m_\sigma = 700$ *MeV,* $f_\pi = 93$ *MeV,* $m_\pi = 139.6$ *MeV, and* $g = 4.84$ *at* $T = 130$ *MeV,* $\mu = 0$.

| Quantity | ESM | LSM | LQM |
| --- | --- | --- | --- |
| Quark Kinetic Energy | 233.847 | 1022.847 | 187.119 |
| Sigma Kinetic Energy | 136.921 | 240.264 | 2107.049 |
| Pion Kinetic Energy | 1.486 | 431.593 | 0.4833 |
| Sigma Interaction Energy | 904.239 | -292.819 | 925.012 |
| Pion Interaction Energy | -11.970 | -1062.938 | -4.848 |
| Meson-Meson Energy | 415.759 | 3281.261 | -62.762 |
| Hedgehog of Mass | 1680.282 | 3620. 2 | 3152.1 |

In the logarithmic sigma model, Table (4) shows that the hedgehog mass is reduced compared to its value in the linear sigma model due to the strong decrease in meson-meson interaction energy. In the extended sigma model [36], the hedgehog mass shows lower values in comparison with the values predicted by the logarithmic sigma model due to the mesonic potential being extended to include higher-order mesonic interactions in the extended sigma model.

The Nambu-Jona-Lasinio (NJL) model was postulated by Nambu and Jona-Lasinio [3]. The NJL model successfully describes nucleon-nucleon interactions by using fermionic local point like interactions in analogy to the gap of BCS theory of superconductors [37]. The model and its extensions successfully study the chiral phase transition of QCD and the meson properties at finite temperature and chemical potential such as in Refs. [38, 39]. The NJL model is not renormalizable therefore one needs the regularization procedure to overcome the divergences integrals. There are several regularization schemes such as cut-off regularization and Pauli-Villars methods are suggested [40].

We consider the NJL model with the midpoint technique which is introduced to calculate meson masses, the constituent quark mass, and the effective potential at finite temperature and chemical potential. In addition, the chiral phase transition and the critical temperature are studied as Ref. [41]. The thermodynamic potential for the NJL model can be written

$$\Omega = \frac{(M-m_0)^2}{2G} - \frac{12}{(2\pi)^3}\int_{-\infty}^{\infty}\{E + T\ln[1+\exp\left(-\frac{(E+\mu)}{T}\right)] +$$

$$T\ln[1+\exp\left(-\frac{(E-\mu)}{T}\right)]\}d^3p. \tag{31}$$

We obtain the analytic expression of $\Omega$ using the N-midpoint method [42]

$$\Omega(T',\mu') = \frac{(M'-m_0')^2}{2G'} - \frac{6}{n\pi^2}\sum_{i=0}^{n-1}\frac{1}{A_i}\left(\ln^2(A_i)\right)\left(\sqrt{M'^2 + \ln^2(A_i)} +\right.$$

$$T'\ln\left(\exp\left(-\frac{1}{T'}\left(\mu' + \sqrt{M'^2 + \ln^2(A_i)}\right)\right) + 1\right) +$$

$$\left.+T'\ln\left(\exp\left(-\frac{1}{T'}\left(-\mu' + \sqrt{M'^2 + \ln^2(A_i)}\right)\right) + 1\right)\right). \tag{32}$$

Therefore, the energy density is defined as follows:

$$E(T',\mu') = -P(T',\mu') + T'\frac{\partial P(T',\mu')}{\partial T'} + \mu'\frac{\partial P(T',\mu')}{\partial \mu'}, \tag{33}$$

Where,

$$P(T',\mu') = -\Omega(T',\mu'). \tag{34}$$

In this paper, the mid-point technique gives a good accuracy in comparison with other models.

**SUMMARY AND CONCLUSION**

In this review article, we review the extended chiral quark models at hot and dense mediums. In particular, we study in detail, the higher-order quark model, logarithmic quark model, quantized quark model, and NJL model which gives qualitatively agreement with other models and lattice QCD.

Recently, the lattice QCD show the critical temperature for chiral symmetry restoration decrease with increasing magnetic field, a phenomenon known a inverse magnetic catalysis. The great majority of effective models fail to explain this phenomenon [43]. Therefore, we need extra effort to develop these models.

**Extended Chiral Quark Models in the Framework of Quantum Chromodynamic: Theory and their Applications in Hot and Dense Mediums**